\documentstyle[epsf]{article}

\author{
Ronald J.Adler and Alexander S. Silbergleit \\
Gravity Probe B, W.W.Hansen Experimental Physics Laboratory,\\ Stanford University, Stanford, CA 94305-4085, USA\\
e-mail: adler@relgyro.stanford.edu; gleit@relgyro.stanford.edu}

\title
{A General Treatment of Orbiting Gyroscope Precession}

\date{\today}

\begin{document}

\maketitle

\begin{abstract}We review the derivation of the metric for a spinning body of any shape and composition using linearized general relativity theory, and also obtain the same metric using a transformation argument. The latter derivation makes it clear that the linearized metric contains only the Eddington $\alpha$ and $\gamma$ parameters, so no new parameter is involved in frame--dragging or Lense--Thirring effects. We then calculate the precession of an orbiting gyroscope in a general weak gravitational field, described by a Newtonian potential (the gravito-electric field) and a vector potential (the gravito-magnetic field). Next we make a multipole analysis of the potentials and the precession equations, giving all of these in terms of the spherical harmonics moments of the density distribution. The analysis is not limited to an axially symmetric source, although the Earth, which is the main application, is very nearly axisymmetric. Finally we analyze the precession in regard to the Gravity Probe B (GP-B) experiment, and find that the effect of the Earth's quadrupole moment ($J_2$) on the geodetic precession is large enough to be measured by GP-B (a previously known result), but the effect on the Lense--Thirring  precession is somewhat beyond the expected GP-B accuracy.
\end{abstract}
\vfill\eject
\section{Introduction}The Gravity Probe B satellite is scheduled to fly in the year 2000~\cite{mac}. It contains a set of gyroscopes intended to test the predictions of general relativity (GR) that a gyroscope in a low circular polar orbit, with  altitude $650\,km$ will precess about 6.6 {\it arcsec/year} in the orbital plane (geodetic precession) and about 42 {\it milliarcsec/year} perpendicular to the orbital plane (LT precession, see ~\cite{mtw}; \cite{or}, secs. 4.7 and 7.8; \cite{wi}, sec. 9.1). In this paper we review the theoretical derivation of these effects and in particular consider the contributions of the Earth's quadrupole and higher multipole fields.

We first review the derivation of the metric for a rotating body using the standard LGRT approach (~\cite{lt}; ~\cite{mtw}; \cite{or}, secs. 4.7 and 7.8). The metric is characterized by a Newtonian scalar potential (the gravito-electric field) and a vector potential (the gravito-magnetic field)~(\cite{or}, sec. 3.5). We then obtain the same result  with a simple transformation argument which clarifies the physical meaning of the metric (\cite{wi}, sec. 4.3). Specifically it makes clear that if the metric of a point mass contains fundamental parameters such as the Eddington parameters $\alpha$  and $\gamma$, then to lowest order the metric of a rotating body contains no {\bf new} fundamental parameters~\cite{nor}. Thus there is no new Lense--Thirring or frame-dragging parameter to be measured by GP-B or any other experiment.

We then derive the precession equations for a gyroscope in a general weak field system, that is for any scalar and vector potential fields~\cite{mtw}. The calculation is valid to first order in the fields and velocities of the source body and the satellite.
The gravitational field of the earth is described by the scalar and vector potentials which depend on the shape of the body and the mass and velocity distribution inside it. We treat both of these fields by a multipole expansion and express the precessions in series of spherical harmonics. We do not limit ourselves with the axially symmetric case which has been thoroughly studied by Teyssandier~(\cite{tey1}),~(\cite{tey2}). In particular we show that for a solid body rotation up to the order $l\leq2$ both precessions depend only on the tensor of inertia of the Earth. 

The major nonspherical contributions to the GP-B precessions are from the Earth's quadrupole moment. The contribution to the geodetic precession which has a magnitude of about 1 part in $10^3$ is detectable by GP-B and quite important for the determination of the parameter $\gamma$ which is to be measured to about 1 part in $10^5$, the most accurate measurement envisioned~(\cite{or}, sec. 3.5 and in particular table 14.2). This contribution has been calculated independently by Wilkins~\cite{wilk} and Barker \& O'Connel~\cite{boc}, and then in the most elegant and general way by Breakwell~\cite{bre}. The contribution to the Lense--Thirring precession is beyond the expected accuracy of GP-B and  close enough to the result by Teyssandier~(\cite{tey1}) obtained from a different Earth's model. 

We also estimate the influence of Moon and Sun to find that only the geodetic contribution of the Sun must be included in the GP-B data reduction, as anticipated.

\section{The Metric in Linearized General Relativity Theory}

We first briefly review this standard derivation of linearized general relativity theory. The metric of a rotating body such as the Earth is obtained by introducing a small perturbation of the Lorentz metric $\eta_{\mu\nu}$, that is 
\begin{equation}
g_{\mu\nu}=\eta_{\mu\nu}+h_{\mu\nu}
\label{eq:1}
\end{equation}
The perturbation $h_{\mu\nu}$ is assumed to be expressed in isotropic space coordinates so that  $h_{11}=h_{22}=h_{33}=h_{s}$. Similarly we suppose the matter producing the metric field is described by the energy--momentum tensor of slow-moving and low density matter with negligible pressure,
\begin{equation}
T^{\mu\nu}=\rho u^{\mu}u^{\nu},
\label{eq:2}
\end{equation}
where $u^{\mu}$ is the 4--velocity and $\rho$ is the matter density. The field equations are:
\begin{equation} 
G_{\mu\nu}=8\pi G\,T_{\mu\nu},
\label{eq:3}
\end{equation}
where $G_{\mu\nu}$ is the Einstein tensor. 

The calculation of $G_{\mu\nu}$ and $T_{\mu\nu}$ to the lowest order in the perturbation is straightforward and results in the following equations:
\begin{equation}
D^2 \left(h_{\mu\nu}-\frac{1}{2}h\eta_{\mu\nu}\right)=16\pi G\,T_{\mu\nu},\qquad h\equiv h^\sigma_\sigma,
\qquad D^2\equiv \frac{\partial^2}{\partial t^2}-\nabla^2;
\label{eq:4}
\end{equation}
\begin{equation}
h^{\mu\nu}_{|\nu}=0
\label{eq:5}
\end{equation}
Here indices are raised and lowered with the Lorentz metric, which is consistent to lowest order, and the slash denotes differentiation. The last expression imposes the so-called Lorentz condition which can always be achieved by a coordinate transformation (to lowest order) and involves no loss of generality; it is also the analog of a gauge choice in electromagnetism. Because we have used isotropic coordinates equation (\ref{eq:4}) leads immediately to $h_{00}=h_{s}$, and the standard classical correspondence gives $h_{00}=h_{s}=2\Phi$. We thereby obtain the wave equations for the scalar and vector potentials:
\begin{equation}
D^2 \Phi=-4\pi G\,\rho
\label{eq:6}
\end{equation}
\begin{equation}
D^2 \vec h=16\pi G\,\rho\vec v,\qquad \vec h=\{h_{01},\,h_{02},\,h_{03}\},
\label{eq:7}
\end{equation}
where $\vec v$ is the velocity of the source. These equations may be solved in the standard way by means of a retarded Green's function. In the case of a time independent system, or a system which changes so slowly that retardation effects may be ignored, the solution is 
\begin{equation}
\Phi(\vec r)=-G\int{\rho(\vec r^{\,\prime})\,d^3\vec r^{\,\prime}\over|\vec r-\vec r^{\,\prime}|}
\label{eq:8}
\end{equation}
\begin{equation}
\vec h(\vec r)=4G\int{\rho(\vec r^{\,\prime})\vec v(\vec r^{\,\prime})\,d^3\vec r^{\,\prime}
\over|\vec r-\vec r^{\prime}|}
\label{eq:9}
\end{equation}
Note that these expressions are analogs of the equations of electrostatics and magnetostatics, which is why one may speak about gravito-electric and gravito-magnetic effects in the linearized theory.

In summary, we may write the (Lense--Thirring) line element as
\begin{equation}
ds^2=\bigl(1+2\Phi\bigr)dt^2-\bigl(1-2\Phi\bigr)d\vec r^{\,2}+2\vec h\cdot d\vec r\,dt
\label{eq:10}
\end{equation}
This is valid to first order in field intensity and source velocity, and will serve as a basis for calculating the gyroscope precession. 

Fig. 1 shows the general aspects of the scalar and vector fields of a spinning body; the vector field generally points in the same direction as the velocity of the surface of the body.

\section{The Metric with Parameters: Derivation by Frame Transformation}

It is, in fact, possible to obtain the above result from a different and physically interesting perspective, and moreover introduce parameters convenient for discussing experimental measurements. Following Eddington, consider the metric of a massive point with a geometric mass $m$ at a large distance $r$, so that $m/r\ll 1$. We expand the Schwarzschild solution in isotropic coordinates for this situation as
$$
ds^2={(1-m/2r)^2\over(1+m/2r)^2}dt^2-(1+m/2r)^4d\vec r^{\,2}
$$
$$
=\biggl(1-{2m\over r}+{2m^2\over r^2}+\dots\biggr)dt^2-
\biggl(1+{2m\over r}+{3m^2\over r^2}+\dots\biggr)d\vec r^{\,2}
$$
Eddington suggested that this be written in terms of parameters as~\cite{edd},~\cite{rob} 
\begin{equation}
ds^2=\biggl(1-\alpha{2m\over r}+\beta{2m^2\over r^2}+\dots\biggr)dt^2-
\biggl(1+\gamma{2m\over r}+\dots\biggr)d\vec r^{\,2},
\label{eq:11}
\end{equation}
where $\alpha$ and $\beta$ and $\gamma$ are equal to $1$ for general relativity. The power series (\ref{eq:11}) is clearly a rather general form for the metric far from a spherical body. 

Since the parameter $m$ which appears in the metric is a constant of integration representing the mass of the central body (specifically $m=GM/c^2$), we may absorb the parameter $\alpha$ into it, which is equivalent to taking $\alpha\equiv1$. This is consistent as long as no independent nongravitational determination of the mass of the body is considered. We will nevertheless retain $\alpha$ in our calculations as a book-keeping device.

Indeed, the parameters in (\ref{eq:11}) may all be viewed as a tool for tracking which terms in the metric contribute to a  gravitational phenomenon such as the perihelion shift of Mercury or the deflection of the starlight by the Sun. Alternatively they may be viewed as numbers which may {\it not} be equal to $1$ if a metric theory other than general relativity is actually valid. In either case they provide a convenient way to express the results of experimental tests of gravity as giving value to the parameters. This parameterized approach has been extended to include many other parameters and has been highly developed under the name parameterized post--Newtonian theory, or PPN ~\cite{wi}. In this paper we take the viewpoint that general relativity is to be tested and emphasize that we are {\it not} using the more general PPN approach. Solar system measurements give $\beta-1=(0.2\pm1.0)\times10^{-3}$ and $\gamma-1=(-1.2\pm1.6)\times10^{-3}$, which are of course entirely consistent with general relativity.

We consider now only phenomena in which the  quadratic term, $\sim m^2/r^2$, in $g_{00}$ is unimportant, that is in which we may ignore $\beta$ and assume that the underlying gravitational theory is linear. Then for a stationary mass,
\begin{equation}
ds^2=\biggl(1-\alpha{2m\over r}\biggr)dt^2-
\biggl(1+\gamma{2m\over r}\biggr)d\vec r^{\,2}\qquad {\rm(stationary\,\,point\,\,mass)}
\label{eq:12}
\end{equation}
Since this is nearly the Lorentz metric, we may generalize it to a moving mass point by simply transforming to a moving system using a transformation that is Lorentzian to the first order in velocity: 
$$
t_r=t-vx,\qquad x_r=x-vt
$$
Here the subscript $r$ labels the system in which the mass is at rest, and which moves at velocity $v$ in the $x$ direction in the laboratory system. This transformation gives the metric for the moving mass point as
$$
ds^2=\biggl(1-\alpha{2m\over r}\biggr)dt^2-
\biggl(1+\gamma{2m\over r}\biggr)d\vec r^{\,2}+(\alpha+\gamma){4m\over r}\,vdx\,dt,
$$
$$ 
{\rm(point\,\,mass\,\,moving\,\,in\,\,}x\,\,{\rm direction})
$$
This obviously generalizes for motion in any direction to
$$
ds^2=\biggl(1-\alpha{2m\over r}\biggr)dt^2-
\biggl(1+\gamma{2m\over r}\biggr)d\vec r^{\,2}+(\alpha+\gamma){4m\over r}\,(\vec v\cdot d\vec r)\,dt
$$
\begin{equation}
{\rm(moving\,\,point\,\,mass)}
\label{eq:13}
\end{equation}
As we assume that our theory is linear to this order (like general relativity), we can superpose the fields of a distribution of such point masses and replace $m/r$ by $\Phi(\vec r)$ from (\ref{eq:8}) and $4m\vec v/r$ by $\vec h(\vec r)$ from  (\ref{eq:9}), resulting in
$$
\Phi(\vec r)=-G\int{\rho(\vec r^{\,\prime})\,d^3\vec r^{\,\prime}
\over|\vec r-\vec r^{\prime}|}
$$
$$
\vec h(\vec r)=4G\int{\rho(\vec r^{\,\prime})\vec v(r^{\,\prime})\,d^3\vec r^{\,\prime}
\over|\vec r-\vec r^{\prime}|}
$$
\medskip
\begin{equation}
ds^2=\bigl(1+ 2\alpha\Phi\bigr)\,dt^2-
\bigl(1-2\gamma\Phi\bigr)\,d\vec r^{\,2}+\bigl(\alpha+\gamma\bigr)\,\bigl(\vec h\cdot d\vec r\bigr)dt
\label{eq:14}
\end{equation}
This agrees with the general relativity result (\ref{eq:12}) when $\alpha=\gamma=1$ but now contains appropriate combinations of Eddington parameters.

We emphasize that this line element has been obtained for any slowly moving mass distribution from the parameterized metric for a stationary mass point by transformation and superposition, and thus no new parameter appears in the expression. Therefore a measurement of a phenomenon which depends on the cross term in (\ref{eq:14}) provides a value for $\alpha+\gamma$  and {\underline not} for some new parameter associated with gravito-magnetism.

 Note also that the result is rather strong since it depends on the observationally verified Schwarzschild metric (\ref{eq:12}), the well-tested Lorenz transformation, and the principle of superposition, valid for any linear theory. 

\section{General Precession Relations}

An orbiting gyroscope has its spin axis parallel displaced in accord with the metric (\ref{eq:14}). We calculate this motion with minimal assumptions about the potentials, which could be the potentials of a nearly spherical and rather uniform rigid body such as the Earth or a potato-shaped body such as in Fig. 1, with some interior mass distribution. We will work always to the first order in the potentials and in the velocities, $\vec v$ and $\vec V$, of the source and orbiting gyroscope. The parallel displacement equation for the gyro spin $S^\mu$ is (~\cite{mtw}; \cite{or}, secs. 4.7 and 7.8; \cite{wi}, sec. 9.1):
\begin{equation}
{dS^\mu\over ds}+
\biggl\{{\mu\atop\nu\sigma}\biggr\}
S^\nu\,{dx^\sigma\over ds}=0
\label{eq:15}
\end{equation}
We suppose that the gyro spin 4-vector is perpendicular to the velocity 4-vector, which is equivalent to assuming that the gyro spin has no zero component in its rest frame. From this assumption the zero component in another frame is easily obtained, and to first order in the satellite velocity it is given by $S^0=\vec S\cdot\vec V$. We calculate the Euler--Lagrange equations in the standard way, and then put them into canonical form to give the Christoffel symbols. To lowest order in the potential and velocity the Christoffel symbols are
$$
\biggl\{{0\atop 0l}\biggr\}=\alpha\Phi_{|l},\quad \biggl\{{i\atop 00}\biggr\}=\alpha\Phi_{|i},\quad 
\biggl\{{i\atop ii}\biggr\}=-\gamma\Phi_{|i}
$$
$$
\biggl\{{i\atop il}\biggr\}=-\gamma\Phi_{|l},\quad\biggl\{{i\atop ll}\biggr\}=\gamma\Phi_{|i},
\quad \biggl\{{i\atop 0l}\biggr\}=\frac{\alpha+\gamma}{4}\,(h_{l|i}-h_{i|l})
$$
The Roman indices in these expressions are space indices and run from 1 to 3, a slash denotes an ordinary derivative, and $l\not=i$. Note that the gravitational vector potential $\vec h$ occurs in a gauge invariant way, that is only its $curl$ appears.

Substitution of the Christoffel symbols into the spin equation of motion gives
\begin{equation}
\dot S^i+(\alpha+\gamma)\Phi_{|i}\left(S^lV^l\right)-\gamma S^i\left(\Phi_{|l}V^l\right)
-\gamma V^i\left(\Phi_{|l}S^l\right)+\left(\frac{\alpha+\gamma}{4}\right)\left(h_{l|i}-h_{i|l}\right)S^l=0
\label{eq:16}
\end{equation}
We break the drift rate into two parts, the geodetic drift rate due to the scalar potential $\Phi$, and the Lense--Thirring precession rate due to the vector potential $\vec h$. Separating also symmetric and antisymmetric parts of the geodetic effect, we arrive at, in a 3--dimensional vector notation,
\begin{equation}
\dot{\vec S}_{LT}=\vec\Omega_{LT}\times\vec S, \qquad \vec\Omega_{LT}\equiv\left({\alpha+\gamma\over4}\right)\nabla\times\vec h,
\label{eq:17}
\end{equation}
and
$$
\dot{\vec S}_{G}=\vec\Omega_{G}\times\vec S+
\biggl\{-
\frac{\alpha}{2}\Bigl[\bigl(\vec S\cdot\vec V\bigr)\nabla\Phi+\bigl(\vec S\cdot\nabla\Phi\bigr)\vec V\Bigr]
+\gamma\bigl(\vec V\cdot\nabla\Phi\bigr)\vec S
\biggr\},
$$
\begin{equation}
\vec\Omega_{G}\equiv\left({\alpha+2\gamma\over2}\right)\nabla\Phi\times\vec V
\label{eq:18}
\end{equation}
\begin{equation}
\dot{\vec S}=\dot{\vec S}_{G}+\dot{\vec S}_{LT},
\label{eq:19}
\end{equation}
where $\vec\Omega_{LT}$ and $\vec\Omega_{G}$ are the instantaneous values  of the Lense--Thirring and geodetic precessions, respectively. Since $\vec\Omega_{LT}$ is the {\it curl} of the gravitational vector potential the Lense--Thirring precession rate is the analog of the magnetic field in magnetostatics theory. 
The geodetic effect is of the order of the scalar potential times the orbital velocity of the satellite, while the Lense--Thirring terms are of the order the scalar potential times the velocity of the central body (the Earth), which is typically much smaller than the orbital velocity. 
Symmetric geodetic terms are responsible for stretching the spin vector $\vec S$, and the reason for separating them is that their effect almost vanishes when averaged over any reasonable satellite orbit. To see this consider the last term in the symmetric part of the geodetic drift rate. Using Newton's law in the form $\nabla\Phi=-\dot{\vec V}$ we may write
$$
\bigl\langle\left(\nabla\Phi\cdot\vec V\right)\bigr\rangle\vec S=-\biggl\langle\left(\frac{d\vec V}{dt}\cdot\vec V\right) \biggr\rangle\vec S
=-\biggl\langle\frac{1}{2}\,\frac{d\vec V^2}{dt}\biggr\rangle\vec S=-\frac{\Delta\vec V^2}{2T}\,\vec S,
$$
where $\Delta\vec V^2$ is the change in the velocity squared in total time $T$, and it is assumed that the drift rate is small and $\vec S$ does not change significantly during this period of time. If the orbit is periodic, this quantity will be zero, and for a nearly periodic orbit it will be very small. Similarly for the remaining symmetric terms in the geodetic drift we may write:
$$
\bigl\langle\bigl(\vec S\cdot\vec V\bigr)\nabla\Phi+\bigl(\vec S\cdot\nabla\Phi\bigr)\vec V\bigr\rangle=
-\biggl\langle\bigl(\vec S\cdot\vec V\bigr)\frac{d\vec V}{dt}+\bigl(\vec S\cdot\frac{d\vec V}{dt}\bigr)\vec V\biggr\rangle=
$$
$$
-\biggl\langle V^i\frac{d V^l}{dt}+V^l\frac{d V^i}{dt}\biggr\rangle S^l=
-\biggl\langle \frac{d V^iV^l}{dt}\biggr\rangle S^l=
-\frac{\Delta V^iV^l}{T}\,S^l
$$
Again, this is zero  for a periodic orbit and very small for a nearly periodic orbit.

In summary the average precession rate of the gyro spin is
\begin{equation}
\langle\dot{\vec S}\rangle=\langle\dot{\vec S}_{G}\rangle+\langle\dot{\vec S}_{LT}\rangle,\qquad 
\langle\dot{\vec S}_{G}\rangle=
\langle\vec\Omega_{G}\rangle\times\vec S,\qquad 
\langle\dot{\vec S}_{LT}\rangle=
\langle\vec\Omega_{LT}\rangle\times\vec S,
\label{eq:20}
\end{equation}
with the values of the geodetic and precessions given in (\ref{eq:17}) and (\ref{eq:18}).

Note that the geodetic and Lense--Thirring effects are approximately perpendicular to each other for an approximately polar orbit around a central body of reasonable shape; for a circular polar orbit about a spherical body they are perpendicular.
 Fig. 2 shows the orientation of the various vectors for a fairly general situation, a satellite in a roughly planar polar orbit about an oddly shaped body. The geodetic precession vector $\vec\Omega_{G}$ is roughly perpendicular to the orbit plane and the Lense--Thirring precession vector $\vec\Omega_{LT}$ has the general appearance of a dipole magnetic field.

\section{Effect of Distant Masses}

If there are relatively distant masses, such as the Moon, in the neighborhood of the central body, their density distribution $\rho_D(\vec r)$ may be expressed as
$$
\rho_D(\vec r)=\sum_{n}M_n\delta(\vec r-\vec r_n)
$$
The effect on the scalar and vector potentials is
$$
\Phi_D(\vec r)=-\sum_{n}{GM_n\over |\vec r-\vec r_n|},\qquad
\vec h_D(\vec r)=4\,\sum_{n}{GM_n\vec v_n\over|\vec r-\vec r_n|}
$$
It then follows that the precessions due to these distant masses are 
\begin{equation}
\vec\Omega^D_{G}=-\frac{\alpha+2\gamma}{2}\,\sum_{n}\frac{GM_n(\vec r-\vec r_n)\times\vec V}{|\vec r-\vec r_n|^3},\qquad
\vec\Omega^D_{LT}=-(\alpha+\gamma)\sum_{n}{G\vec L_n\over|\vec r-\vec r_n|^3},
\label{eq:21}
\end{equation}
where $\vec L_n=M_n(\vec r-\vec r_n)\times\vec v_n$ is the angular momentum of the distant mass.

For the specific case of the Moon and the GP-B experiment the numerical values will be discussed in sec. 9.

\section{Multipole Expansions}

We will henceforth study the case of a body such as the Earth which is rigidly rotating, that is $\vec v=\vec\omega\times\vec r$. Motivated by expression (\ref{eq:9}) we introduce a new vector potential quantity $\vec\Pi(\vec r)$:
\begin{equation}
\vec h(\vec r)=4\vec\omega\times\vec\Pi(\vec r),\qquad 
\vec\Pi(\vec r)=G\int{\rho(\vec r^{\,\prime})\vec r^{\,\prime}\,d^3\vec r^{\,\prime}
\over|\vec r-\vec r^{\prime}|}
\label{eq:22}
\end{equation}
The vector $\vec\Pi(\vec r)$ is a harmonic function outside the body which we will expand in spherical harmonics; for a special spherically symmetric case $\vec\Pi(\vec r)$ is collinear with $\vec r$. In terms of $\vec\Pi(\vec r)$, the metric (\ref{eq:14}) may be written as
\begin{equation}
ds^2=\bigl(1+ 2\alpha\Phi\bigr)\,dt^2-
\bigl(1-2\gamma\Phi\bigr)\,d\vec r^{\,2}+4\bigl(\alpha+\gamma\bigr)\,\bigl(\vec\omega\times\vec\Pi\bigr)\cdot d\vec rdt
\label{eq:23}
\end{equation}

Using (\ref{eq:8}) and (\ref{eq:9}) it is possible to express the divergence of $\vec\Pi(\vec r)$ via the scalar potential, namely,
$$
\nabla\cdot\vec\Pi=-(\Phi+\vec r\cdot\nabla\Phi)
$$
which allows us to rewrite formula (\ref{eq:17}) for the Lense--Thirring precession as
\begin{equation}
\vec\Omega_{LT}=
-(\alpha+\gamma)\,\biggl[(\vec\omega\cdot\nabla)\vec\Pi+\vec\omega\Bigl(\Phi+\vec r\cdot\nabla\Phi\Bigr)\biggr]
\vec\Omega_{LT}=
-(\alpha+\gamma)\omega\,\biggl[\vec\Pi_{|z}+\Bigl(\Phi+r\Phi_{|r}\Bigr)\biggr]\hat z 
\label{eq:24}
\end{equation}
Here we have chosen the $z$ axis to be along the spin, $\vec\omega=\omega\,\hat z$.

We now put the origin of spherical coordinates $\{r,\theta,\varphi\}$ at the center of mass of the body, introduce spherical harmonics with the notation
\begin{equation}
Y^{\nu}_{lm}(\theta,\varphi)=P_l^m(\cos\theta)f^\nu(m\varphi), \qquad
f^\nu(m\varphi)=\cases{\cos m\varphi,\,\,\nu=c\cr\sin m\varphi,\,\,\nu=s\cr},
\label{eq:25}
\end{equation}
and expand the potentials $\Phi$ and $\vec\Pi$ in corresponding series:
\begin{equation}
\Phi(\vec r,t)=-{GM\over r}\Biggl[1+\sum_{l\geq2,\,m,\,\nu}a^{\nu}_{lm}(t)\biggl({R\over r}\biggr)^lY^{\nu}_{lm}(\theta,\varphi)\Biggr]
\label{eq:26}
\end{equation}
\begin{equation}
\Pi_i(\vec r,t)=\,{GMR\over r}\Biggl[\sum_{l\geq1,\,m,\,\nu}p^{i\nu}_{lm}(t)\biggl({R\over r}\biggr)^lY^{\nu}_{lm}(\theta,\varphi)\Biggr]
\label{eq:27}
\end{equation}
[$l=1$ terms in (\ref{eq:26}) and $l=0$ terms in (\ref{eq:27}) are missing because the origin is at the center of mass of the body, see expressions (\ref{eq:28}) and (\ref{eq:29}) below].
These expansions are in the inertial frame in which the body rotates; the direction of the coordinate polar axis z is so far arbitrary; $R$ is the characteristic size of the body, which we take to be the equatorial radius for the earth. The potentials are slowly varying functions of time due to the earth rotation, so we write the coefficients as explicit functions of time. These coefficients are related to the mass distribution in the standard way by
\begin{equation}
a^{\nu}_{lm}(t)={(2-\delta_{m0})\over M}\,{(l-m)!\over(l+m)!}\,
\int\,\rho(\vec r,t)\,\biggl({r\over R}\biggr)^lY^{\nu}_{lm}(\theta,\varphi)\,d^3\vec r 
\label{eq:28}
\end{equation}
\begin{equation} 
p^{i\nu}_{lm}(t)={(2-\delta_{m0})\over M}\,{(l-m)!\over(l+m)!}\,
\int\,\rho(\vec r,t)\,\biggl({r\over R}\biggr)^l\,{x_i\over R}\,Y^{\nu}_{lm}(\theta,\varphi)\,d^3\vec r
\label{eq:29}
\end{equation}
Note that generally the domain of integration here is also time dependent. In particular, we write for convenience $a^{c}_{l0}(t)\equiv a_{l0}(t),\,\, a^{s}_{l0}(t)\equiv 0\,\, p^{ic}_{l0}(t)\equiv p^{i}_{l0}(t),\,\, p^{is}_{l0}(t)\equiv0$.
 
The time dependence in (\ref{eq:28}) and (\ref{eq:29}) may be easily analyzed for a rigidly rotating body provided that the rotation axis is fixed both in the inertial space and in the body. In that case in the inertial frame with the $z$ axis along the spin, $\vec\omega=\omega\,\hat z$, the earth density is
$$
\rho(\vec r,t)=\rho_e(r,\theta,\varphi-\omega t)\equiv\rho_e(r,\theta,\bar\varphi),
$$
where $\rho_e(r,\theta,\bar\varphi)\equiv\rho_e(\vec r)$ is the time independent density measured in the frame rotating with the earth. Substituting this for the density in (\ref{eq:28}) and (\ref{eq:29}) and transforming to the earth fixed frame we find that the time dependent coefficients may be written in terms of constant moments of density according to
\begin{equation}
\left[\matrix{a^{c}_{lm}(t)\cr a^{s}_{lm}(t)\cr}\right]=
\left|\left|\matrix{\cos\omega t&-\sin\omega t\cr\sin\omega t&\cos\omega t \cr}\right|\right|
\left[\matrix{a^{c}_{lm}\cr a^{s}_{lm}\cr}\right]
\label{eq:30}
\end{equation}
\begin{equation}
\left[\matrix{p^{ic}_{lm}(t)\cr p^{is}_{lm}(t)\cr}\right]=
\left|\left|\matrix{\cos\omega t&-\sin\omega t\cr\sin\omega t&\cos\omega t \cr}\right|\right|
\left[\matrix{p^{ic}_{lm}\cr p^{is}_{lm}\cr}\right],
\label{eq:31}
\end{equation}
and the time independent coefficients are explicitly given by
\begin{equation}
a^{\nu}_{lm}=a^{\nu}_{lm}(0)={(2-\delta_{m0})\over M}\,{(l-m)!\over(l+m)!}\,
\int\,\rho_e(\vec r)\,\biggl({r\over R}\biggr)^lY^{\nu}_{lm}(\theta,\bar\varphi)\,d^3\vec r 
\label{eq:32}
\end{equation}
\begin{equation} 
p^{i\nu}_{lm}=p^{i\nu}_{lm}(0)={(2-\delta_{m0})\over M}\,{(l-m)!\over(l+m)!}\,
\int\,\rho_e(\vec r)\,\biggl({r\over R}\biggr)^l\,{x_i\over R}\,Y^{\nu}_{lm}(\theta,\bar\varphi)\,d^3\vec r
\label{eq:33}
\end{equation}

If the rotation axis wanders in the body and/or in the inertial space, the relation between the time dependent and time independent coefficients is given by a combination of appropriate rotation matrices which is more complicated than the one matrix in (\ref{eq:30}) and (\ref{eq:31}) (see c. f.~\cite{ro}). Generally, a time dependent coefficient with the indices $l$ and $m$ is a linear combination of the appropriate time independent coefficients with the indices $l$ and $n=0,1,\dots,l$.

Introducing now the time dependent coefficients from (\ref{eq:32}) and (\ref{eq:33}) back into (\ref{eq:26}) and (\ref{eq:27}) we obtain the time dependent potentials in a convenient form using time independent coefficients only:
\begin{equation}
\Phi(\vec r,t)=-{GM\over r}\Biggl[1+\sum_{l\geq2,\,m,\,\nu}a^{\nu}_{lm}\biggl({R\over r}\biggr)^lY^{\nu}_{lm}(\theta,\varphi-\omega t)\Biggr]
\label{eq:34}
\end{equation}
\begin{equation}
\Pi_i(\vec r,t)=\,{GMR\over r}\Biggl[\sum_{l\geq1,\,m,\,\nu}p^{i\nu}_{lm}\biggl({R\over r}\biggr)^lY^{\nu}_{lm}(\theta,\varphi-\omega t)\Biggr]
\label{eq:35}
\end{equation}
This is of course what one should expect intuitively; in general this form will be most useful for our purposes. 

The constant coefficients $a^{\nu}_{lm}$ in (\ref{eq:34}) are those that are measured very accurately for the Earth (up to $l=18$ their values are found in~\cite{dma}).  However, for the earth of arbitrary shape and composition it is impossible to express $p^{i\nu}_{lm}$ through $a^{\nu}_{lm}$, in other words, their values, and hence the scalar and vector potentials, are independent. Nevertheless, a useful relationship between the two sets of coefficients exists; to describe it, we need a notation for a general moment of the density,
\begin{equation}
M^{\nu}_{klm}\equiv\int\,\rho(\vec r)\,\biggl({r\over R}\biggr)^kY^{\nu}_{lm}(\theta,\varphi)\,d^3\vec r;
\label{eq:36}
\end{equation}
in particular,
\begin{equation}
a^{\nu}_{lm}={(2-\delta_{m0})\over M}\,{(l-m)!\over(l+m)!}M^{\nu}_{llm}
\label{eq:37}
\end{equation}
Using the definitions (\ref{eq:32}) and (\ref{eq:33}) and the recurrence relations for Legendre functions~\cite{bate}, we derive the following equalities relating $p^{i\nu}_{lm}$ to $a^{\nu}_{lm}$: 
$$
p^{1\nu}_{lm}=(2l+1)^{-1}\Bigl\{
-2^{-1}(l+m+1)(l+m+2)a^{\nu}_{l+1m+1}+(2-\delta_{m1})^{-1}a^{\nu}_{l+1m-1}+
$$
$$
\Bigl[(l-m)!/M(l+m)!\Bigr]
\left[M^{\nu}_{l+1l-1m+1}-(l+m-1)(l+m)M^{\nu}_{l+1l-1m-1}\right]
\Bigr\}
$$
$$
p^{2\nu}_{lm}=(\mp)(2l+1)^{-1}\Bigl\{
2^{-1}(l+m+1)(l+m+2)a^{\mu}_{l+1m+1}+(2-\delta_{m1})^{-1}a^{\mu}_{l+1m-1}-
$$
\begin{equation}
\Bigl[(l-m)!/M(l+m)!\Bigr]
\left[M^{\mu}_{l+1l-1m+1}-(l+m-1)(l+m)M^{\mu}_{l+1l-1m-1}\right]
\Bigr\}
\label{eq:38}
\end{equation}
$$
p^{3\nu}_{lm}=
(2l+1)^{-1}\Bigl\{
(l+m+1)a^{\nu}_{l+1m+1}+
(2-\delta_{m0})(l-m)!M^{\nu}_{l+1l-1m}
/M(l+m-1)!
\Bigr\}
$$
In the second line of (\ref{eq:38}), the minus sign is taken and $\mu=s$ when $\nu=c$, the plus sign and $\mu=c$ when $\nu=s$.

From (\ref{eq:34}) and (\ref{eq:35}) using the definition (\ref{eq:17}) of $\vec\Omega_{LT}$ a rather compact multipole expansion for the Lense--Thirring precession may be computed:
\begin{equation}
\Omega_{LT}^{i}(\vec r,t)=\frac{\alpha+\gamma}{2}{GM\omega\over r}\,
\sum_{l\geq2,\,m,\,\nu}\Bigl[(l-m)\,p^{i\nu}_{l-\delta_{i3}\,m}-l\,a^{\nu}_{lm}\delta_{i3}\Bigr]\biggl({R\over r}\biggr)^l
Y^{\nu}_{lm}(\theta,\varphi-\omega t)
\label{eq:39}
\end{equation}
The corresponding expansion for the geodetic precession is too cumbersome to be useful.

\section{The Far Field/High Symmetry Approximation: Multipoles with $l\leq2$ and the Tensor of Inertia}

If the shape of the central body and the mass distribution inside it are known, then all the pertinent quantities, including multipole expansion coefficients $a^{\nu}_{lm}$ and $p^{i\nu}_{lm}$, may be found by integration, but this is rarely the case. Even when $a^{\nu}_{lm}$ are measured, as for the Earth, all the $p^{i\nu}_{lm}$, and the Lense--Thirring effect with them, remain entirely undetermined. However, for a body of any shape and composition, the coefficients $a^{\nu}_{2m},\,(l=2)$ and $p^{i\nu}_{1m},\,(l=1)$ can be expressed in terms of the elements $I_{ij}$ of the tensor of inertia ${\bf I}$ determined in a standard way,
$$
I_{ij}=\int\,\rho(\vec r)\,( r^{2}\delta_{ij}-x_ix_j)\,d^3\vec r
$$
Writing $I_{ii}\equiv I_i$, we find:
$$
a_{20} = -\frac{2I_{3}-I_{2}-I_{1}}{2MR^2},
\quad a^{c}_{22}=\frac{I_{2}-I_{1}}{4MR^2},
\quad p^{3}_{10}=-\frac{I_{3}-I_{2}-I_{1}}{2MR^2};
$$
\begin{equation}p^{1c}_{11}=-\frac{I_{3}+I_{2}-I_{1}}{2MR^2}
\qquad p^{2s}_{11}=-\frac{I_{3}-I_{2}+I_{1}}{2MR^2};
\label{eq:40}
\end{equation}
$$
a^{c}_{21}=-p^{1}_{10}=p^{3c}_{11}=\frac{I_{13}}{MR^2},
\qquad a^{s}_{21}=-p^{2}_{10}=p^{3s}_{11}=\frac{I_{23}}{MR^2},
$$
$$
 -a^{s}_{22}=\,0.5\,{p^{1s}_{11}}\,=\,0.5\,{p^{2c}_{11}}\,=\frac{I_{12}}{MR^2}
$$
This is done by comparing the integrals (\ref{eq:32}) (with $l=2$) and (\ref{eq:33}) (with $l=1$) to $I_{ij}$ using explicit expressions of Legendre functions with $l=1,2$; formulas for $a_{20}$ and $a^{c}_{22}$ are known and used in geodesy for the determination of the Earth's moments of inertia~\cite{bu}. 

Introducing (\ref{eq:40}) into (\ref{eq:34}), (\ref{eq:35}) and dropping the terms with $l>2$ for $\Phi$ and $l>1$ for $\vec\Pi$, we first obtain the $l\leq2$ formulas for the potentials:
\begin{equation}
\Phi(\vec r)=-{G\over r}
\Biggl[
M+{1\over2r^2}\,\biggl({\rm tr}\,{\bf I}-{3\over r^2}\,\bigl({\bf I}\vec r\cdot\vec r\bigr)\biggr)
\Biggr],\quad
\vec\Pi(\vec r)=-{G\over r^3}
\Biggl[
{\bf I}\vec r-{1\over2}\,\bigl({\rm tr}\,{\bf I}\bigr)\vec r
\Biggr]
\label{eq:41}
\end{equation}
From these we then obtain the approximation for the precessions by differentiation of the expressions (\ref{eq:18}) and (\ref{eq:17}) (see also (\ref{eq:22})): 
\begin{equation}
\vec\Omega_{G}={(\alpha+2\gamma)G\over 2r^3}
\Biggl\{
\biggl[M+{3\over2r^2}\,\Bigl({\rm tr}\,{\bf I}-{5\over r^2}\,\bigl({\bf I}\vec r\cdot\vec r\bigr)\Bigr)\biggr]
\bigl(\vec r\times\vec V\bigr)+
{3\over r^2}\,\bigl({\bf I}\vec r\times\vec V\bigr)
\Biggr\}
\label{eq:42}
\end{equation}
\begin{equation}
\vec\Omega_{LT}={(\alpha+\gamma)G\over r^3}
\Biggl\{
{\bf I}\vec\omega-3\,\biggl[{1\over2}\,\bigl({\rm tr}\,{\bf I}\bigr)-{1\over r^2}\,\bigl({\bf I}\vec r\cdot\vec r\bigr)\biggr]\vec\omega+
3\,{\vec\omega\cdot\vec r\over r^2}\,\biggl[{1\over2}\,\bigl({\rm tr}\,{\bf I}\bigr)\vec r-{\bf I}\vec r\biggr]
\Biggr\}
\label{eq:43}
\end{equation}
[Of course, the same expression for $\vec\Omega_{LT}$ is also obtained from (\ref{eq:40}) and (\ref{eq:39}) with $l=2$].

Note that formulas (\ref{eq:40}) and (\ref{eq:41}) hold in both the body-fixed and inertial frames, and the coefficients and moments of inertia in the latter depend on the time if the rotation is not axisymmetric, $a^{\nu}_{lm}=a^{\nu}_{lm}(t)$, $p^{i\nu}_{lm}=p^{i\nu}_{lm}(t)$, $I_{ij}=I_{ij}(t)$. The expressions (\ref{eq:42}) and (\ref{eq:43}) for the precessions are meaningful in the inertial frame where generally ${\bf I}={\bf I}(t)={\bf R}(t){\bf I}(0){\bf R}^T(t)$, ${\bf R}(t)$ being a rotation matrix which converts the body-fixed radius-vector into the inertial one. In the above case of a simple rotation about an inertially-fixed axis, $\vec\omega=\omega\hat z$,
$$
{\bf R}(t)=\left|\left|\matrix{
\cos\omega t&-\sin\omega t&0\cr
\sin\omega t&\cos\omega t&0 \cr
0&0&1\cr
}\right|\right|
$$

The results (\ref{eq:41})---(\ref{eq:43}) are valid under either of the two conditions: 
\smallskip

1) {\it far field}, $R/r\ll1$; 

2) {\it high symmetry}, i.e., all higher order moments are small 
\smallskip

\noindent These results alter somewhat our usual notion of the geodetic and Lense--Thirring effects: the first one is generally proportional not only to the orbital momentum, but to ${\bf I}\vec r\times\vec V$ as well, and the Lense-Thirring precession generally points not only in the direction of the angular momentum $\vec L={\bf I}\vec\omega$, but has also components parallel to $\vec\omega,\,\vec r$, and ${\bf I}\vec r$. Two particular cases of the inertia tensor  are of special interest.

{\it a) Spherical symmetry}, ${\bf I}={\rm diag}\,\{{I},\,{I},\,{I}\}$. In this case, the standard formulas follow immediately from (\ref{eq:43}) for $\alpha=\gamma=1$:
\begin{equation}
\vec\Omega_{G}={3GM\over 2r^3}\,\bigl(\vec r\times\vec V\bigr),\qquad\vec
\Omega_{LT}=
{2GI\over r^3}\,
\biggl[
-\vec\omega+{3\over r^2}\,(\vec\omega\cdot\vec r)\vec r
\biggr]
\label{eq:44}
\end{equation}
Note that we have thus shown this to be the exact result for a spherical earth with any {\it radial} density distribution $\rho=\rho(r)$.

{\it b) Symmetric top,} ${\bf I}={\rm diag}\,\{{I_1},\,{I_1},\,{I}\}$, ${I_1}\not={I}$. It turns out that for $\vec\omega=\omega\hat z$, i. e., for the rotation about the material symmetry axis, the previous expression for $\Omega_{LT}$ remains true; to lowest order in the oblateness, this also proves to be the exact result for a slightly oblate uniform ellipsoid of revolution rotating about its semiminor axis. The corresponding expression for $\Omega_{G}$ is given and discussed in sec. 9.

\section{Earth Models}

To go beyond the $l\leq2$ approximation, one must make some assumptions about the shape of the central body and density distribution inside it, and use the data of gravitational potential measurements, as available. Bearing in mind the application to GP-B, that is, the Earth, we use the following set of assumptions.
\medskip

{\it 1. Gravitational potential.} We assume it to be measured, i.e., the gravitational coefficients $a_{lm}^\nu$ known. Our problem is thus to determine the vector potential in terms of the coefficients $p_{lm}^{i\nu}$ using as general an earth model as possible.
\medskip

{\it 2. Shape.} We assume that the earth is a slightly oblate ellipsoid of revolution, so its surface equation to lowest order in the eccentricity $\epsilon$ is 
\begin{equation}
r=r_s(\theta,\varphi)=R(1-\epsilon\cos^2\theta)
\label{eq:45}
\end{equation}
where $R$ is the semi-major axis (equatorial radius); the eccentricity is the ratio of the difference of the semi-axes to the major one.
\medskip

{\it 3. Mass distribution.} We examine two different models:
\smallskip

{\it a)} With $\rho_0>0$ and $\Delta\rho$ being arbitrary functions of their arguments, we set
\begin{equation}
\rho(\vec r)=\rho_0(r)+\Delta\rho(\theta,\varphi)>0,\quad 
\int\limits_{\rm sphere}\Delta\rho(\theta,\varphi)\sin\theta\,d\theta d\varphi=0
\label{eq:46}
\end{equation}
The first term here describes any depth variation of the {\it average} density, and the only assumption is that the angular variations are depth--independent.
\smallskip

{\it b)} For arbitrary functions $\rho_0>0$ and $\rho_s$, we set
\begin{equation}
\rho(\vec r)=\rho_0(r)+\rho_s(\theta,\varphi)\delta(r-r_s(\theta,\varphi))>0
\label{eq:47}
\end{equation}
Contrary to the previous model, here all angular variations of the density are concentrated at the earth's surface. 
\medskip

We will call model A the set of assumptions $1,\,2$ and $3a)$, and model B the assumptions $1,\,2$ and $3b)$; in both cases the mass distributions are, of course, assumed consistent with the gravitational data of the assumption $1$.

As far as the Earth is concerned, assumption 2) reflects the classical Clairaut formula (see c. f.~\cite{roy}) (the eccentricity is $\epsilon\approx3.353\times10^{-3}$); also the mass model $3b)$ seems rather realistic for the Earth since the estimated thickness of the layer where its mass distribution varies significantly in the angular directions is only about $30\,km$.

Note that an entirely different Earth model is used in geodesy: it assumes that the sum of gravitational and centrifugal potentials is constant at the Earth's ellipsoid's surface~(\cite{ka}). This allows one to relate $a_{l0}$ to the eccentricity and the Earth angular velocity only (in particular, to obtain the Earth's gravitational  oblateness $J_2=-a_{20}$ with a surprisingly good accuracy), but gives zero values to $a^\nu_{lm},\,\, m\not=0$ and leaves $p^{i\nu}_{lm}$ undetermined.

Evidently, our two models should give bounds for the corrections to the $l\geq2$ values of the precessions. Moreover, for both of them it is possible to find the corrections explicitly by means of the following four steps: 
\medskip

$1^\circ$ calculate $a_{lm}^\nu$ by (\ref{eq:32}) via spherical harmonics coefficients $\rho_{lm}^\nu$ of the function $\Delta\rho(\theta,\varphi)$ for model A, [or $\rho_s(\theta,\varphi)$ for model B]; 
\medskip

$2^\circ$ since $a_{lm}^\nu$ are assumed known, solve the resulting equations for $\rho_{lm}^\nu$, hence having them expressed via  $a_{lm}^\nu$ (thus $\Delta\rho$ or $\rho_s$ are uniquely determined at this stage through the gravitational data);
\medskip

$3^\circ$ using that, calculate the needed moments $M^\nu_{l+1l-1m}$ of the density by (\ref{eq:36}) in terms of $a_{lm}^\nu$;
\medskip

$4^\circ$ using the found values of $M^\nu_{l+1l-1m}$, express $p_{lm}^{i\nu}$ through $a_{lm}^\nu$ according to (\ref{eq:38}).
\medskip 

In fact, what is described here is a fit of our density distributions (\ref{eq:46}) and (\ref{eq:47})
to the known gravitational coefficients $a_{lm}^\nu$ which allows us, in these two cases, to express the former through the latter and thus determine uniquely the coefficients $p_{lm}^{i\nu}$, i.e., the gravito-magnetic part of the field. The implementation of this procedure is rather cumbersome, though basically straightforward, so the details are given in the Appendix. The results of the calculations to lowest order in the Earth's eccentricity are:
$$
p^{1\nu}_{lm}={1\over 2(2l+1)}\,
\Biggl[
-{(l+m+1)(l+m+2)}a^\nu_{l+1m+1}+
{(l-m)(l-m-1)\kappa_l}a^\nu_{l-1m+1}+
$$
$$
{2\over2-\delta_{m1}}\biggl(a^\nu_{l+1m-1}-\kappa_l a^\nu_{l-1m-1}\biggr)
\Biggr]+O\bigl((\epsilon l)^2\bigr)
$$
$$
p^{2\nu}_{lm}={(\mp)\over 2(2l+1)}\,
\Biggl[
{(l+m+1)(l+m+2)}a^{\mu}_{l+1m+1}+
{(l-m)(l-m-1)\kappa_l}a^{\mu}_{l-1m+1}-
$$
\begin{equation}
{2\over2-\delta_{m1}}\biggl(a^{\mu}_{l+1m-1}+\kappa_la^{\mu}_{l-1m-1}\biggr)
\Biggr]+O\bigl((\epsilon l)^2\bigr)
\label{eq:48}
\end{equation}
$$
p^{3\nu}_{lm}={1\over 2l+1}\,
\Biggl[
(l+m+1)a^\nu_{l+1m}+
(l-m)\kappa_la^\nu_{l-1m}
\Biggr]+
O\bigl((\epsilon l)^2\bigr),
$$
with the minus sign and $\mu=s$ for $\nu=c$, plus sign and $\mu=c$ for $\nu=s$ in the second of these formulas, and
$$
\kappa_l={l+2\over l+4},\quad {\rm model\,\,A}\qquad \kappa_l=1,\quad {\rm model\,\,B}
$$

It is remarkable that the first non-zero oblateness correction is quadratic, and that the results for the two models differ by just a factor $(l+2/l+4)$ in front of a couple of terms. These expressions are needed only for $l\geq2$ since for $l=1$ formulas (\ref{eq:40}) always hold.

\section{Results for Gravity Probe B}

We now apply the above results to the GP-B satellite which will be circling the Earth on a low ($R/r\approx0.9$) polar nearly circular orbit. Our aim is to check the GR predictions for the geodetic and Lense-Thirring drift of a GP-B gyroscope taking into account all the peculiarities of the real Earth's gravito-electric and gravito-magnetic field pertinent to the expected experimental error of one part in $10^{5}$ for the geodetic effect and about one part in $10^{2}$ for the Lense-Thirring effect. More precisely, one needs to check whether the spherically symmetric approximation (\ref{eq:44}), which has been used for many years in the theoretical discussion and planning of the experiment, needs any corrections to match the above experimental accuracy. 

We use the GR values for the Eddington parameters $\alpha=\gamma=1$ in the sequel. We always use the assumption 2 of the previous section that the Earth has a shape of a slightly oblate ellipsoid of revolution with the eccentricity $\epsilon=3.353\times10^{-3}$. Its semi-minor axis is assumed fixed in the inertial space and coincident with the Earth rotation axis and the $z$ axis of the Cartesian coordinates.
\medskip

 {\it 1. Gravitational potential}. For the required accuracy, it is enough to include only the Earth's quadrupole moment into $\Phi$, i. e., to set
\begin{equation}
-a_{20}\equiv J_2\approx1.083\times10^{-3};\qquad
a^\nu_{lm}=0\quad{\rm for}\quad l=2,\,m>0; \quad l>2,
\label{eq:49}
\end{equation}
This is because all gravitational coefficients other than the Earth's oblateness $J_2$ are at least 2 orders of magnitude smaller~\cite{dma}. By (\ref{eq:49}) and (\ref{eq:40}), the $l\leq2$ expression (\ref{eq:40}) for $\Phi$ is valid with 
\begin{equation}
{\bf I}={\rm diag}\,\{{I_1},\,{I_1},\,{I}\},\qquad I_1=I-J_2MR^2,
\label{eq:50}
\end{equation}
which gives the gravitational potential in the usual form:
$$
\Phi(\vec r)=-\frac{GM}{r}\left[1-J_2\frac{R^2}{2r^2}\left(3\frac{z^2}{r^2}-1\right)\right]=
-\frac{GM}{r}\left[1-J_2\frac{R^2}{r^2}P_2(\cos\theta)\right]
$$
\medskip

{\it 2. Instantaneous geodetic precession}. Just as for the potential, (\ref{eq:49}) assures that the $l\leq2$ formula (\ref{eq:42}) is valid for $\Omega_G$. The explicit form of the geodetic precession is obtained either from (\ref{eq:42}) and (\ref{eq:50}) or from its general definition (\ref{eq:18}) by differentiating the above expression for the potential. The result reads:
\begin{equation}
\vec\Omega_{G}={3GM\over 2r^3}\,
\left\{\left(\vec r\times\vec V\right)+J_2\frac{3}{2}\left(\frac{R}{r}\right)^2
\left[
\left(1-5\frac{z^2}{r^2}\right)\left(\vec r\times\vec V\right)+2z\left(\hat z\times\vec V\right)
\right]
\right\},
\label{eq:51}
\end{equation}
This expression contains the $J_2$ corrections to the classical spherically symmetric expression (\ref{eq:44}). Breakwell~\cite{bre} obtained it but in a  different form, pinning the value of the precession to a point on a given satellite orbit rather than to a given point in space through which the spinning particle passes with a velocity $\vec V$. 
\medskip

{\it 3. Gravito-magnetic field and the instantaneous Lense--Thirring precession}. We now invoke our assumption about the mass distribution (models A and B of the previous section) and substitute the values (\ref{eq:49}) of the gravitational coefficients into the expressions (\ref{eq:48}) for $p^{i\nu}_{lm}$. In this way it turns out that for {\it both} models A and B the only additional to $l=1$ non-zero coefficients are those with $l=3,\,\,m=0,1$, and their values are
$$
p^{1c}_{31}=p^{2s}_{31}=\frac{5}{49}\,J_2,\quad  p^{3}_{30}=-\frac{15}{49}\,J_2,\qquad{\rm model\,\,A};
$$
\begin{equation}
p^{1c}_{31}=p^{2s}_{31}=\frac{1}{7}\,J_2,\quad  p^{3}_{30}=-\frac{3}{7}\,J_2,\qquad{\rm model\,\,B};
\label{eq:52}
\end{equation}
So, for both our models the only addition to the field of a symmetric top is the gravito-magnetic field (\ref{eq:52}). In terms of the Lense--Thirring precession that means that the standard expression (\ref{eq:44}) has to be appended by the terms (\ref{eq:52}) from the general formula (\ref{eq:39}) for the precession. The result reads:
$$
\vec\Omega_{LT}={GI\omega\over r^3}\left[
\left(3\frac{z^2}{r^2}-1\right)\hat z+3\frac{z}{r}{x\hat x+y\hat y\over r}\right]-
$$
\begin{equation}
J_2\frac{GMR^2\omega}{r^3}\,\frac{R}{r}
\left[Z\left(\frac{z^2}{r^2}\right)\frac{R}{r}\hat z+X\left(\frac{z^2}{r^2}\right){x\hat x+y\hat y\over r}\right],
\label{eq:53}
\end{equation}
where (recall $z^2/r^2=\cos^2\theta$)
\begin{equation}
Z(\cos^2\theta)=Z_0\,(35\cos^4\theta-30\cos^2\theta+3),\qquad 
X(\cos^2\theta)=X_0\,(5\cos^2\theta-1);
\label{eq:54}
\end{equation}
$$
Z_0=\frac{15}{98},\quad  X_0=\frac{15}{49},\quad{\rm model\,\,A};\qquad
Z_0=\frac{3}{14},\quad  X_0=\frac{3}{7},\quad{\rm model\,\,B}
$$
The second line of (\ref{eq:53}) is the correction to the classical Lense--Thirring expression in the first line which is induced by the gravito-magnetic field (\ref{eq:52}). [Recall from section 7, b), that the dependence of $\vec\Omega_{LT}$ on $J_2$ for a symmetric top rotating about its symmetry axis cancels out - remarkably!. Thus the first line of (\ref{eq:53}) contains no $J_2$]. 

Note that the only difference between the Earth's models A and B is a factor of 7/5 in the constants $Z_0$ and $X_0$. 

Note also that allowing for more multipoles in the gravitational potential than given in the equation (\ref{eq:49}) we automatically obtain higher order multipoles in the gravito-magnetic field by the expressions (\ref{eq:48}) for $p^{i\nu}_{lm}$. However, for GP-B all these higher order corrections prove to be too small to be taken into account.
\medskip

{\it 4. Orbit of the GP-B satellite and the instantaneous values of precessions}. The perfect orbit for the GP-B satellite would be a circular polar one with the altitude $h_s=650\, km$ which is described by $r=r_0=R+h_s,\qquad V=V_0=\sqrt{{GM}/{r_0}}$. In reality it is slightly distorted by the quadrupole moment of the Earth's gravitational field; with the lowest order in $J_2$ corrections included, the orbit becomes~(\cite{bre2}),~(\cite{boc}):
\begin{equation}
r=r_0\,
\left[
1-{1\over4}J_2\left({R\over r_0}\right)^2\cos2\theta
\right],
\label{eq:55}
\end{equation}
$$
\vec V =V_{0}\,
\left[
1-{3\over8}\,J_2\,\left({R\over r_0}\right)^2
\right]
\left[
\hat {\theta}+{1\over2}\,J_2\left({R\over r_0}\right)^2\sin2\theta\,\hat {r} 
\right],
$$
where $\hat {r}, \hat {\theta}, \hat {\varphi}$ are the corresponding unit vectors. (In view of the orbit symmetry, it is enough to consider a half-orbit from one pole to the other, $0\leq\theta\leq\pi$; otherwise, the second half may be assigned to $\pi\leq\theta\leq2\pi$, with the proper direction of the unit vector $\hat e_{\theta}$).

Introducing the orbital velocity  (\ref{eq:55}) into the formula (\ref{eq:51}) and dropping some terms $O(J_2^2)$,  we obtain the geodetic precession for the GP-B gyroscope in the following form:
$$
\vec\Omega_{G}={3GMV_{0}\over 2r^2}
\left[
1-{3\over8}\,J_2\,\left({R\over r_0}\right)^2
\right]
\Biggl[
1+{3\over2}\,J_2\biggl({R\over r_0}\biggr)^2\,\biggl(1-3\cos^2\theta\biggr)
\Biggr]\hat e_{\varphi} 
$$
Using now the orbit radius (\ref{eq:55}) and keeping only the corrections linear in $J_2$, we arrive at the expression we need:
\begin{equation}
\vec\Omega_{G}={3GMV_{0}\over 2r_0^2}
\Biggl[
1+J_2\biggl({R\over r_0}\biggr)^2\,\left(\frac{5}{8}-\frac{7}{2}\cos^2\theta\right)
\Biggr]\hat e_{\varphi}
\label{eq:56}
\end{equation}

To derive a similar formula for $\vec\Omega_{LT}$ from (\ref{eq:53}) we, of course, need no orbital velocity but only the orbit radius (\ref{eq:55}); with accuracy $O(J_2)$ the instantaneous value of the Lense--Thirring precession for GP-B turns out to be
\vfill\eject
$$
\vec\Omega_{LT}=\vec\Omega_{LT}^{(0)}+\vec\Omega_{LT}^{(J_2)};
$$
\smallskip
\begin{equation}
\vec\Omega_{LT}^{(0)}=\frac{GI\omega}{r_0^3}
\left[\left({3\cos^2\theta}-1\right)\hat z+\left(3\cos\theta\sin\theta\right)\hat x\right];
\label{eq:57}
\end{equation}
\smallskip
$$
\vec\Omega_{LT}^{(J_2)}=J_2\frac{GI\omega}{r_0^3}\left[
\left(\frac{R}{r_0}\right)^2\left(U-Z\frac{MR^2}{I}\right)\hat z+
\left(V\frac{R}{r_0}\cos\theta-X\frac{MR^2}{I}\right)\sin\theta\hat x
\right]
$$
For convenience, we have assumed that the orbit is in the plane $y=0$; the values of $Z=Z\left(\cos^2\theta\right)$ and $X=X\left(\cos^2\theta\right)$ are given in (\ref{eq:54}), while
\begin{equation}
U(\cos^2\theta)=\frac{3}{4}(6\cos^4\theta-5\cos^2\theta+1),\qquad 
V(\cos^2\theta)=\frac{9}{4}(2\cos^2\theta-1)
\label{eq:58}
\end{equation}

Note that the expressions (\ref{eq:56}) and (\ref{eq:57}) contain the corrections coming both directly from the non-spherically symmetric gravito-electric and  gravito-magnetic field and also through the influence of the former on the orbit. The first contribution can be recognized from the second in the Lense--Thirring precession (\ref{eq:57}) by the presence of the factor $MR^2/I$, however, in the geodetic precession (\ref{eq:56}) this difference is concealed.
\medskip

{\it 5. Orbit-averaged precessions: the GP-B drift rates}. With the above results, it is now straightforward to carry out the orbit averaging which reduces to a simple integration over $\theta$. In this way, to lowest order in the Earth's oblateness $J_2\approx1.083\times10^{-3}$ using (\ref{eq:56}) we find
\begin{equation}
\biggl\langle\vec\Omega_{G}\biggr\rangle={3GMV\over 2r_0^2}\,
\Biggl[
1-{9\over8}\,J_2\,\biggl({R\over r_0}\biggr)^2
\Biggr]\hat e_{\varphi}={3GMV\over 2r_0^2}\,
\Biggl[
1-1.003\times10^{-3}
\Biggr]\hat e_{\varphi},
\label{eq:59}
\end{equation}
where the values $R=6,378\,km$ and $r_0=7,028\,km$ for the Earth's and orbit radius have been taken. The result exactly coincides with the one found by Breakwell~\cite{bre} in a different way (he gave the formula for the orbit of any inclination). Since GP-B is intended to measure the geodetic precession and the Eddington parameter $\gamma$ to about a part in $10^5$, a  $0.1\%$ correction is critically important. Note that the GP-B gyro spin axis will be initially aligned with the reference direction to the Guide Star in the orbit plane, so the geodetic effect will cause the in-plane drift of the spin in the North--South direction.

In a similar fashion, from (\ref{eq:57}) we derive
\begin{equation}
\biggl\langle\vec\Omega_{LT}\biggr\rangle={GI\omega\over 2r^3_0}\,
\Biggl[
1+{9\over4}\,J_2\,\biggl({R\over r_0}\biggr)^2\biggl(\frac{1}{2}-Z_0{MR^2\over I}\biggr)
\Biggr]\hat z
\label{eq:60}
\end{equation}
For our model B, $Z_0=3/14$ according to  (\ref{eq:54}); therefore, with ${MR^2/I}=3.024$ for the Earth,
\begin{equation}
\biggl\langle\vec\Omega_{LT}\biggr\rangle={GI\omega\over 2r^3_0}\,
\Biggl[
1+{9\over8}\,J_2\,\biggl({R\over r_0}\biggr)^2\biggl(1-\frac{3}{7}{MR^2\over I}\biggr)
\Biggr]\hat z=
{GI\omega\over 2r^3_0}\,
\Biggl[
1-2.97\times10^{-4}
\Biggr]\hat z
\label{eq:61}
\end{equation}
Thus the $J_2$ correction for the model B is $(-0.03\%)$, which is essentially beyond the expected GP-B accuracy for the Lense--Thirring effect. For the model A the correction is of the opposite sign and almost one order of magnitude smaller, about $0.007\%$. For a different model of the mass distribution inside the Earth Teyssandier~\cite{tey1} has obtained a slightly larger correction of about $(-0.011\%)$, with the same relevance to GP-B.
\medskip

{\it 6. Effect of the Moon}. According to formula (\ref{eq:21}) from section 5 the geodetic effect from a distant mass such as the Moon scales with the mass and inversely with the square of distance. Since the Moon has a small mass and is at a large distance we expect its effect to be small. A rough estimate for the geodetic precession due to the Moon is
\begin{equation}
\frac{\Omega_G^M}{\bigl\langle\Omega_{G}\bigr\rangle}=\frac{M_M}{M}\left(\frac{r_0}{r_M}\right)^2\approx
\,\,10^{-6}
\label{eq:62}
\end{equation}
which is too small to be of significance for GP-B experiment.

The Lense--Thirring effect scales with the angular momentum and inversely with the cube of the distance. Since the velocity of the Moon in its orbit is small we expect its effect to be small, and the estimate
\begin{equation}
\frac{\Omega_{LT}^M}{\bigl\langle\Omega_{LT}\bigr\rangle}=\frac{L_M}{L}\left(\frac{r_0}{r_M}\right)^3\approx
\,\,10^{-5}
\label{eq:63}
\end{equation}
shows that it is really far beyond the GP-B accuracy. Thus the moon is of no consequence for the GP-B experiment.
\medskip

{\it 6. Effect of the Sun}. Although the Sun is quite distant its mass is large compared to the Earth so it is not apparent how large its effect on the precession will be. We can calculate this using (\ref{eq:18}). In the inertial frame centered on the Sun the velocity of the satellite will be the sum of the satellite velocity in the Earth orbit plus the orbital velocity of the Earth
\begin{equation}
\vec V=\vec V_S+\vec V_{EO}
\label{eq:64}
\end{equation}
But the satellite velocity averages to zero in the course of one orbit, while the other factors in (18) change very little, so for long time averages we may neglect the satellite velocity compared to the Earth orbital velocity. Treating the Sun as a point mass and approximating the Earth orbit as a circle we may thus write (\ref{eq:18}) as
\begin{equation}
\vec\Omega_{G}^S\approx\left({\alpha+2\gamma\over2}\right)\,\frac{GM_S V_{EO}}{r_S^2}\,\hat n,
\label{eq:65}
\end{equation}
where the unit vector $\hat n$  is perpendicular to the plane of the ecliptic and $r_S$ is the Earth-Sun distance. The numerical value of this, with $\alpha=\gamma=1$, is
\begin{equation}
\bigl\langle\Omega_{G}^S\bigl\rangle\approx6.3\times 10^{-7}rad/yr=19\,\,marcsec/yr
\label{eq:66}
\end{equation}
This contribution to the GP-B precession is not negligible and must be included in the data analysis. It was first discussed by deSitter in 1916~\cite{deS1},~\cite{deS2}, and further information and references can be found in the book of Will~\cite{wi}.
        Note an important fact concerning the direction of the above precession vector: it is perpendicular to the plane of the ecliptic, whereas the geodetic precession vector due to the Earth lies mainly in the equatorial plane of the Earth, so the two are not parallel. Indeed the precession due to the Sun is roughly in the direction of the Lense-Thirring precession due to the Earth.
        Finally we may estimate the precession due to the spin of the Sun by using (\ref{eq:44}). Since the spin period of the Sun is about 1 month this gives roughly
\begin{equation}
\bigl\langle\Omega_{LT}^S\bigl\rangle\approx 10^{-11}\,rad/yr
\label{eq:67}
\end{equation}
which is far too small to be of relevance to the GP-B experiment.

\section*{Acknowledgments}
 
This work was supported by NASA grant NAS 8-39225 to Gravity Probe~B. We are grateful to the members of GP-B Theory Group, especially, to C.W.F.Everitt, G.M.Keiser, R.V.Wagoner, and P.W.Worden, for many fruitful discussions and enlightening comments. 

\section*{Appendix}

Let us show briefly the implementation of the four-step procedure described in section 8 which allows us to obtain expressions (\ref{eq:48}) for gravito-magnetic multipole coefficients $p^{i\nu}_{lm}$. We do it for the Earth's model A; calculations for the model B are similar. 

First, we introduce the model A density distribution (\ref{eq:46}) into the definition (\ref{eq:36}) of the general moment $M^\nu_{klm}$ of the density and calculate this moment for the shape of a slightly oblate ellipsoid of revolution (assumption 2 of section 8). Working to the first order in the eccentricity $\epsilon$, we find $M^\nu_{klm}$ in terms of the spherical harmonics coefficients
$$
\rho^\nu_{lm}=\int\Delta\rho Y^{\nu}_{lm}\sin\theta d\theta d\varphi
$$
of the function $\Delta\rho(\theta,\varphi)$ from (\ref{eq:46}), namely:
$$
R^{-3}M^\nu_{klm}=\frac{1-\epsilon(k+3)Q_{lm}}{k+3}\rho^\nu_{lm}-
\epsilon(k+3)[S_{lm}\rho^\nu_{l+2,m}+T_{lm}\rho^\nu_{l-2,m}]+O\bigl((\epsilon k)^2\bigr)
\eqno(A.1)
$$
Here $Q_{lm},\,S_{lm},\,T_{lm}$ are known positive rational fractions of $l$ and $m$ bounded for all pertinent values of those parameters,
$$
Q_{lm}=\frac{(l-m+1)(l+m+1)}{(2l+1)(2l+3)}+\frac{(l+m)(l-m)}{(2l-1)(2l+1)},
$$
$$
S_{lm}=\frac{(l-m+1)(l-m+2)}{(2l+1)(2l+3)},
\qquad T_{lm}=\frac{(l+m)(l+m-1)}{(2l-1)(2l+1)};
$$
Formula (A.1) is slightly different for the case $l=m=0$, but we do not need it. For $k=l$ the left-hand side of the equality (A.1) is given via $a^\nu_{lm}$ according to (\ref{eq:37}):
$$
R^{-3}\frac{M}{2-\delta_{m0}}\frac{(l+m)!}{(l-m)!}a^\nu_{lm}=
$$
$$
\frac{1-\epsilon(l+3)Q_{lm}}{l+3}\rho^\nu_{lm}-
\epsilon(l+3)[S_{lm}\rho^\nu_{l+2,m}+T_{lm}\rho^\nu_{l-2,m}]+O\bigl((\epsilon l)^2\bigr)
$$
This is a tri-diagonal system of linear algebraic equations for $\rho^\nu_{lm}$ with small off-diagonal elements. Solving it for $\rho^\nu_{lm}$, we express the latter in terms of $a^\nu_{lm}$:
$$
\frac{R^{3}}{M}\rho^\nu_{lm}=\frac{l+3}{N_{lm}}a^\nu_{lm}+
\eqno(A.2)
$$
$$
\epsilon\left[
\frac{l+3}{N_{lm}}Q_{lm}a^\nu_{lm}+\frac{l+5}{N_{l+2m}}S_{lm}a^\nu_{l+2m}+\frac{l+1}{N_{l-2m}}T_{lm}a^\nu_{l-2m}
\right]+O\bigl((\epsilon l)^2\bigr),
$$
where 
$$
N_{lm}=\left(2-\delta_{m0}\right)\frac{(l-m)!}{(l+m)!}
$$
Introducing now the expression (A.2) back into the general formula (A.1), we get {\it all} moments, with {\it any k}, expressed through the gravitational coefficients:
$$
M^\nu_{klm}=\frac{M}{2-\delta_{m0}}
\biggl\{
\frac{(l+m)!}{(l-m)!}\frac{l+3}{k+3}a^\nu_{lm}-
\eqno(A.3)
$$
$$
\epsilon \frac{k+2}{k+3}\Bigl[\tilde V_{lm}a^\nu_{lm}+\tilde S_{lm}a^\nu_{l+2,m}+\tilde T_{lm}a^\nu_{l-2,m}\Bigr]
\biggr\}
+O\bigl((\epsilon l)^2\bigr), 
$$
with the quantities $\tilde Q_{lm},\,\tilde S_{lm},\,\tilde T_{lm}$ simply related to $Q_{lm},\,S_{lm},\,T_{lm}$, respectively:
$$
\tilde Q_{lm}=\frac{(l+m)!}{(l-m)!} Q_{lm},\quad \tilde S_{lm}=\frac{(l+m+2)!}{(l-m+2)!} S_{lm},\quad
\tilde T_{lm}=\frac{(l+m-2)!}{(l-m-2)!} T_{lm}
$$
We finally set $k=l\pm1$ in (A.3) and use the resulting expressions in the relation (\ref{eq:38}), completing thus the last fourth step of the procedure described in section 8 and obtaining the equalities (\ref{eq:48}) for $p^{i\nu}_{lm}$ (model A).

\vfill\eject
\centerline{\epsfxsize=6in\epsffile{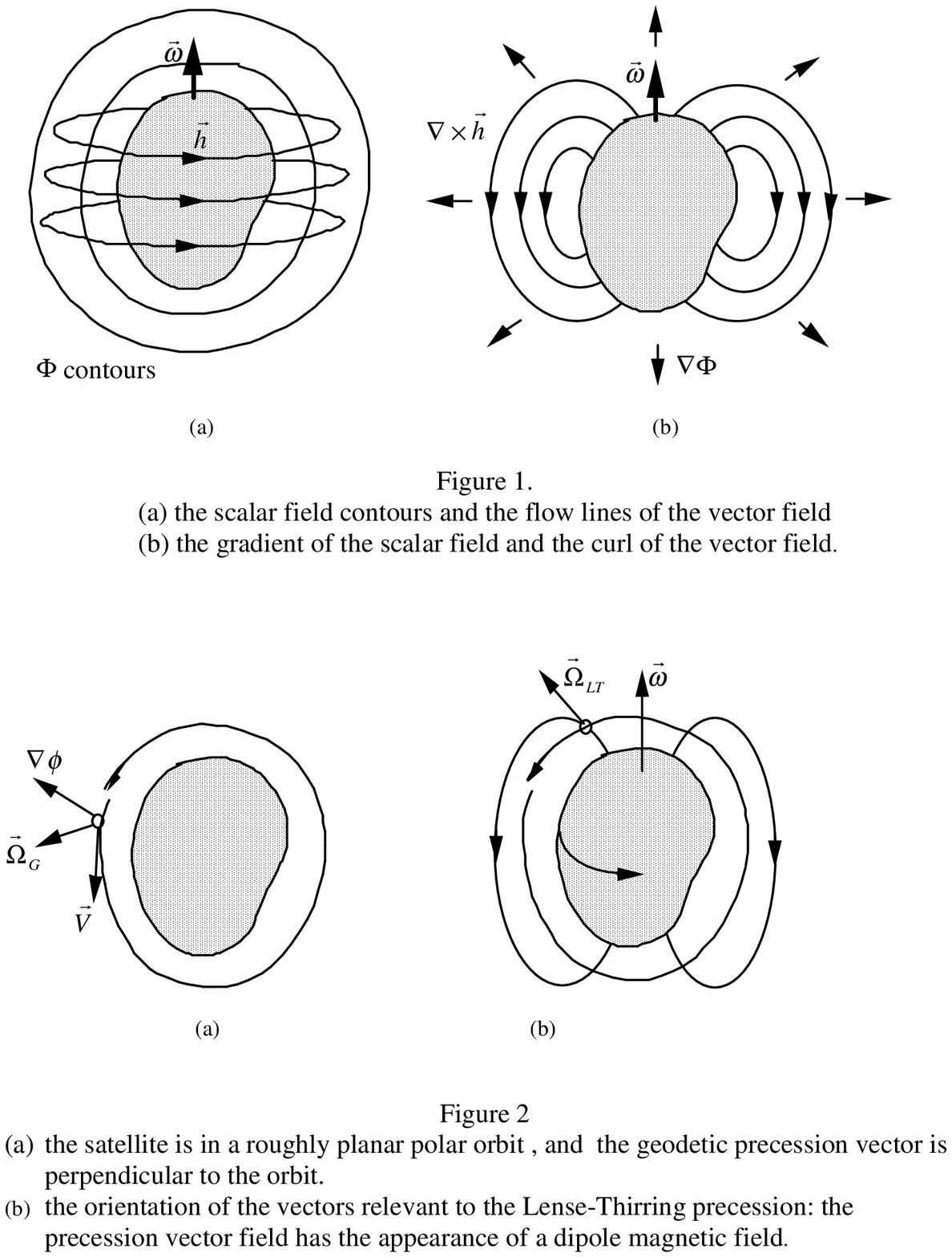}}
\vfill\eject

\end{document}